\documentclass[sigplan, authorversion=true, nonacm=true]{acmart}
\usepackage{graphicx} 
\usepackage{booktabs}
\usepackage{listings}
\usepackage{color}
\usepackage{xcolor}
\usepackage{algorithm}
\usepackage[noend]{algpseudocode}
\usepackage{varwidth}
\lstdefinestyle{mycodestyle}{
    backgroundcolor=\color{white},   
    commentstyle=\color{teal},
    keywordstyle=\color{blue},
    numberstyle=\tiny\color{gray},
    stringstyle=\color{purple},
    basicstyle=\ttfamily\footnotesize,
    breakatwhitespace=false,         
    breaklines=true,                 
    captionpos=b,                    
    keepspaces=true,                                 
    showspaces=false,                
    showstringspaces=false,
    showtabs=false,                  
    tabsize=2
}
\newcommand{\vecgpt}{\textsc{LLM-Vectorizer}}
\begin{document}
\title{LLM-Vectorizer: LLM-based Verified Loop Vectorizer}

\author{Jubi Taneja}
\affiliation{
  \institution{Microsoft Research, USA}
  \country{}
}
\email{jubitaneja@microsoft.com}
\author{Avery Laird}
\authornote{Work done while interning at Microsoft Research, USA}
\affiliation{
  \institution{University of Toronto, Canada}
  \country{}
}
\email{alaird@cs.toronto.edu}
\author{Cong Yan}
\affiliation{
  \institution{Microsoft Research, USA}
  \country{}
}
\email{congyan.me@gmail.com}

\author{Madan Musuvathi}
\affiliation{
  \institution{Microsoft Research, USA}
  \country{}
}
\email{madanm@microsoft.com}
\author{Shuvendu K. Lahiri}
\affiliation{
  \institution{Microsoft Research, USA}
  \country{}
}
\email{shuvendu.lahiri@microsoft.com}

\begin{abstract}
Vectorization is a powerful optimization technique that significantly boosts the performance of high performance computing applications operating on large data arrays. Despite decades of research on auto-vectorization, compilers frequently miss opportunities to vectorize code. 
On the other hand, writing vectorized code manually using compiler intrinsics is still a complex, error-prone task that demands deep knowledge of specific architecture and compilers.

In this paper, we evaluate the potential of large-language models (LLMs) to generate vectorized (Single Instruction Multiple Data) code from scalar programs that process individual array elements. 
We propose a novel finite-state-machine  {\it multi-agents} based approach that harnesses LLMs and test-based feedback to generate vectorized code.
Our findings indicate that LLMs are capable of producing high-performance vectorized code with run-time speedup ranging from $1.1x$ to $9.4x$ as compared to the state-of-the-art compilers such as Intel Compiler, GCC, and Clang.

To verify the correctness of vectorized code, we use Alive2, a leading bounded translation validation tool for LLVM IR.
We describe a few domain-specific techniques to improve the scalability of Alive2 on our benchmark dataset. Overall, our approach is able to verify $38.2\%$ of vectorizations as correct on the TSVC benchmark dataset. 
\end{abstract}
\maketitle
\section{Introduction}
In today's data-driven world, loop vectorization plays a crucial role in accelerating the high performance computing (HPC) and AI applications. Vectorization allows operations to be performed on entire arrays simultaneously, which is much faster than iterating over elements one-by-one. To this end, many decades of research effort has been invested in automatic vectorization. 
However, compilers frequently fail to apply such optimizations due to the imprecision of static analysis.
Vectorization is driven by data dependence analysis which is difficult to establish precisely due to complexities of code including complex control flow, aliasing, discontinuous memory access \cite{feng2021evaluation, maleki2011evaluation,siso2019evaluating,madan2021auto}.
Further, compilers use conservative cost models to determine the profitabilty of vectorization that often leads to poor choices of optimization~\cite{maleki2011evaluation}.

To enable users to exploit vectorization opportunities directly, frameworks such as AVX2~\cite{avx, avx2} introduce a set of compiler {\it intrinsics} (C-level functions) that allow invoking the assembly level single instruction multiple data (SIMD) instructions from the C source code.
%
For example, the intrinsic \texttt{\_mm256\_loadu\_si256}  allows for the loading of $256$ bits of integer data from memory into a YMM register,
while \texttt{\_mm256\_storeu\_si256} enables the storage of $256$ bits from a YMM register back to memory. 
These intrinsics, among others, empower developers to write code that is both performant and maintainable.
However, this puts the burden of ensuring the correctness of the vectorized program on the user. 
Given the subtle semantics of these operations, writing such programs is complex, error-prone and is limited to expert users only. 

Recent advances in large-language models (LLMs) have demonstrated the potential to generate and transform code based on natural language instructions~\cite{chen2021evaluating, transcoder}.
On the other hand, formal verification for LLVM has matured with the advent of tools such as Alive2~\cite{lopes2021alive2}.
Motivated by these progress, we ask the following question to democratize the creation of vectorized programs by non-experts:
\begin{quote}
\textit{Can advances in LLMs and formal verification be leveraged to automatically optimize scalar C programs into equivalent vectorized programs using AVX2 intrinsics?}
\end{quote}
\newcommand{\myrule}{\vskip0.15in\hrule\vskip0.10in}
\begin{figure*}
{\small
%
\myrule
\begin{minipage}[t]{0.40\textwidth}
\begin{lstlisting}[language=C]
void s212(int n, int *a, int *b, int *c, int *d) {
    for (int i = 0; i < n-1; i++) {
        a[i] *= c[i];
        b[i] += a[i + 1] * d[i];
    }
}
\end{lstlisting}
(a) Unvectorized C code
\myrule
\vskip 0.2in
\includegraphics[scale=0.50]{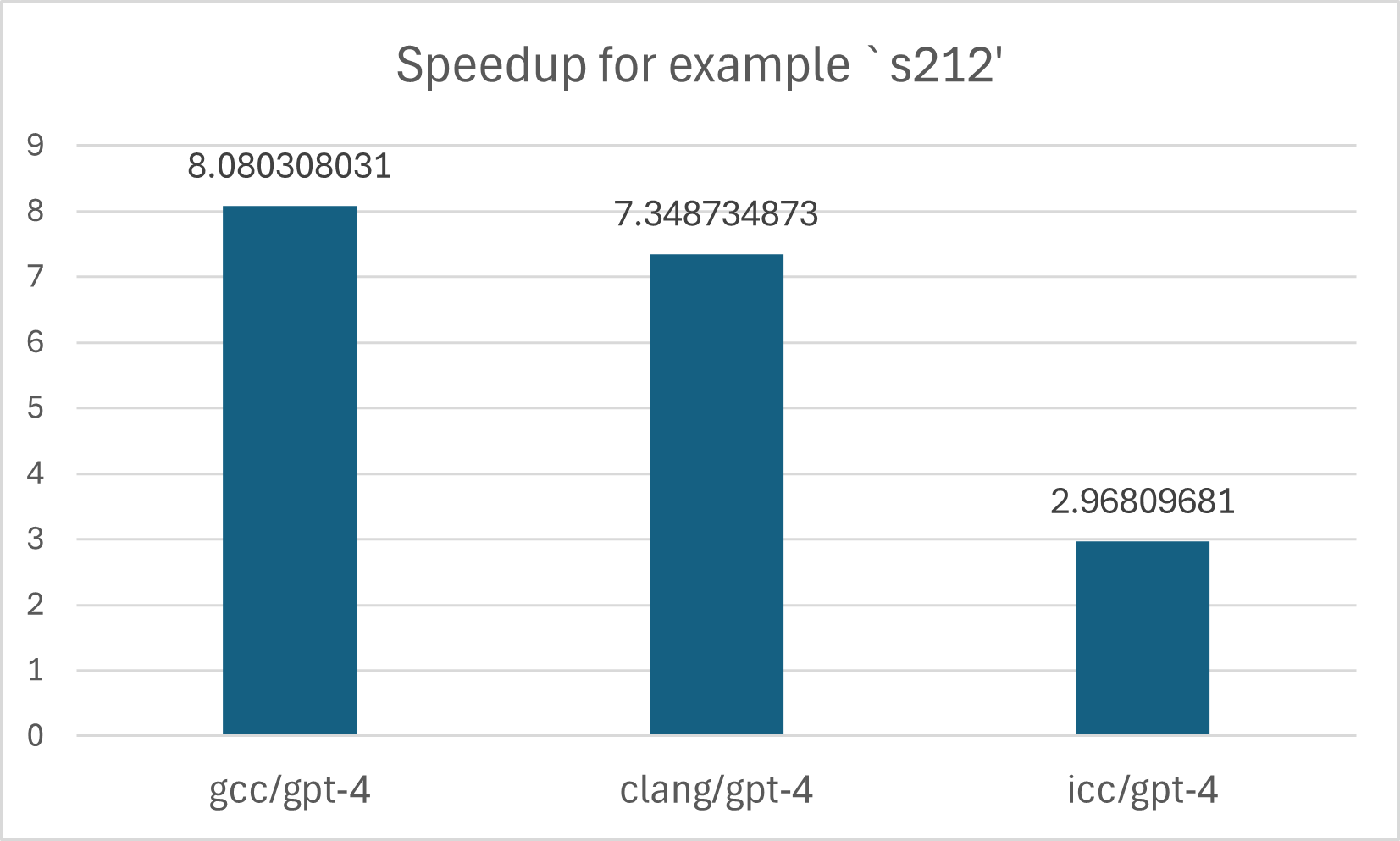}
\vskip 0.2in
(c) Runtime Performance Speedup
\label{fig:speedup-chart}
\end{minipage}
%
%
\hspace{0.035\textwidth}
\begin{minipage}[t]{0.55\textwidth}
\begin{lstlisting}[language=C]
#include <immintrin.h>

void s212(int n, int *a, int *b, int *c, int *d) {
  int i;
  __m256i a_vec, b_vec, c_vec, a_next_vec, d_vec, prod_vec, sum_vec;
  for (i = 0; i < n-1 - (n-1)%8; i+=8) {
    a_vec = _mm256_loadu_si256((__m256i *) &a[i]);
    b_vec = _mm256_loadu_si256((__m256i *) &b[i]);
    c_vec = _mm256_loadu_si256((__m256i *) &c[i]);
    a_next_vec = _mm256_loadu_si256((__m256i *) &a[i+1]);
    d_vec = _mm256_loadu_si256((__m256i *) &d[i]);
    prod_vec = _mm256_mullo_epi32(a_vec, c_vec);
    _mm256_storeu_si256((__m256i *) &a[i], prod_vec);
    prod_vec = _mm256_mullo_epi32(a_next_vec, d_vec);
    sum_vec = _mm256_add_epi32(b_vec, prod_vec);
    _mm256_storeu_si256((__m256i *) &b[i], sum_vec);
  }
  for (; i < n-1; i++){
    a[i] *= c[i];
    b[i] += a[i + 1] * d[i];
  }
}
\end{lstlisting}
(b) Vectorized C Code generated by GPT-4
\end{minipage}
\myrule
}
\caption{An example from TSVC benchmark that cannot be vectorized by the state-of-the-art compilers (GCC, Clang, ICC) but by GPT-4.}
\label{fig:motivation-perf}
\end{figure*}
%

%

%
In Figure \ref{fig:motivation-perf}, a motivating example is presented from the TSVC benchmark \cite{maleki2011evaluation}. The input C function, as illustrated in Figure \ref{fig:motivation-perf}(a), represents a scalar program, and is not vectorized by state-of-the-art compilers, including GCC, Clang, and ICC. The prevalent cause for this nonvectorization is attributed to the compilers' detection of a backward loop-carried dependency that occurs when accessing the array element $a[i+1]$ concurrently with the operation of writing to $a[i]$. This perceived dependency hinders the compilers' ability to apply vectorization to the code segment.
However, this dependence is spurious.
In the scalar code, the operation of writing to $a[i]$ depends on read operation of current elements of arrays, $a$ and $c$ at index $i$.
The operation of writing to $b[i]$ depends on read operation of the subsequent element in the array $a$, denoted as $a[i+1]$, current elements of arrays $b$ and $d$.
The update to $b[i]$ is specifically based on the preceding value of $a[i+1]$, which has not been modified by the first statement in the loop. This indicates the non-existence of cross-dependence that could interfere with the vectorization optimization.

Upon prompting GPT-4~\cite{achiam2023gpt} to generate vectorized code targeting AVX2, we obtain the C code utilizing the intrinsics of AVX2, as shown in Figure \ref{fig:motivation-perf}(b).
 Intutitively, intrinsics such as \texttt{\_mm256\_mullo\_epi32} perform element-wise multiplication of $32$-bit integers within a $256$-bit register, producing lower $32$ bits of the result.
 Likewise, intrinsic \texttt{\_mm256\_add\_epi32} perform element-wise addition of eight $32$-bit integers within a $256$-bit register.

GPT-4  produces the vectorized code by loading two different vectors of array $a$. The vector $a\_vec$ incorporates eight integer elements starting from the index $i^{th}$ and another vector $a\_next\_vec$ includes eight integer elements starting from the index $(i+1)^{th}$.
Subsequently, arithmetic operations are performed on these vectors in parallel, and the results are stored back into the arrays $a$ and $b$.

We compile the vectorized code in Figure \ref{fig:motivation-perf}(b) with Clang without any additional optimizations to study the effect of the LLM-based vectorized code in isolation. On compilation, LLM-based vector code shows a speedup of $2.09x$, $7.35x$, and $8.08x$ as compared to Icc, Clang and GCC compilers respectively as shown in 
Figure \ref{fig:motivation-perf}(c).

Although LLMs exhibit {\it creativity} in the generation of high-performance vectorized code (in this case), it does not provide any assurance of correctness of the generated code. 
%
%
We introduce \vecgpt{}, an end-to-end tool that leverages
LLMs to automatically vectorize code correctly.
By ingesting natural language prompts augmented with a scalar C code snippet, \vecgpt{} generates corresponding vectorized C code.
Internally, our design harnesses recent work on {\it AI-based agents}~\cite{wu2023autogen} and orchestrate the interaction between such agents (some of whom may be LLMs in turn) using a finite state machine (FSM). This FSM is designed to reduce the number of LLM invocations, and to facilitate the repair of incorrect vectorized code generated by LLM through a feedback loop. The multiple agents rely on a simple checksum-based testing as the correctness criterion.

In this paper, we improve the trust on the correctness of the plausible (possibly correct) vectorized code generated by \vecgpt{}, by subjecting it to {\it bounded translation validation} (a form of symbolic formal refinement checking as implemented in Alive2~\cite{lopes2021alive2}).
We demonstrate that the verification queries often causes Alive2 to time-out; we provide techniques to scale the verification exploiting domain-specific insights for the case of the vectorization problem.

%
\paragraph{Overview} We summarize the results of our exploration with \vecgpt{} and Alive2 below. 
\begin{itemize}
    \item We find that LLMs are capable of producing efficient vectorized code, often outperforming modern compilers that take a cautious approach to transformation in order to ensure soundness guarantees. (Section \ref{sec:results-perf}).
    \item More concretely, \vecgpt{} demonstrates the capability to generate plausibly correct vectorized code for $125$ out of $149$ test programs in the Test Suite for Vectorizing Compilers (TSVC) benchmark~\cite{maleki2011evaluation}, as per the checksum-based testing criterion (Section \ref{sec:design}, \ref{sec:results-csum}).
    \item The use of symbolic verification tool (through Alive2) increases the number of vectorizations that can be proven inequivalent by $2.5x$ (from $24$ to $61$) compared to checksum-based testing. Of these, our domain-specific optimizations contribute to proving $20$ more vectorizations incorrect. 
    \item Of the $125$ plausible cases, we have used Alive2 and our domain-specific optimizations to formally verify (modulo loop unrolling) the correctness for $57$ ($38.2\%$) of the test programs. Once again, our optimizations contribute to the verification of $31$ additional examples where Alive2 timed out.
    \item Of the $125$ plausible vectorizations, $31$ remain inconclusive due to various reasons, including timeouts, out-of-memory issues, unmodeled functions and encoding of vector intrinsics in Alive2. (Section \ref{sec:results-equivalence})
    \item The multiple agents FSM design in \vecgpt{} identifies $24$ new test programs that can be vectorized with just one LLM invocation and also successfully repairs inequivalent vectorized code using a feedback loop. (Section \ref{sec:fsm})
    \item We categorize common causes where LLMs encounter challenges in vectorizing programs including loop-carried dependencies and unsafe hoisting. (Section \ref{sec:results-failed})
\end{itemize}
\paragraph{Contributions}
In summary, the paper makes the following contributions:
\begin{enumerate}
    \item We evaluate the capabilities of LLMs for the task of performing vectorization through a source-to-source transformation. 
    \item We describe a novel {\it AI-agents}-based approach to provide LLMs with feedback to improve the quality of suggestions. 
    \item We employ formal verification techniques (namely bounded translation validation) to assess the correctness of the transformation at the LLVM level.
    \item We provide domain-specific optimizations that help in scaling Alive2 to both verify as well as refute almost twice as many examples. 
\end{enumerate}
\paragraph{Organization}
The rest of the paper is organized as follows: Section~\ref{sec:design} provides the description of \vecgpt{} including the use of AI-agents. Section~\ref{sec:equivalence} describes the symbolic verification using Alive2 as well as our domain-specific optimizations to simplify the task for Alive2. Section~\ref{sec:results} discusses the various dimensions for evaluating our approach with different research questions. We discuss related works in Section~\ref{sec:related} and conclude in Section~\ref{sec:conclusions}.

\section{Design Overview}
\label{sec:design}
\vecgpt{} is an end-to-end tool that uses LLMs to
auto-vectorize code correctly.
This tool initiates its process by ingesting a prompt provided by the user in natural language, along with a scalar program code snippet written in C. The prompt simply asks for a vectorized program for an AVX2 target for a given input unvectorized program.
Upon receiving the input, \vecgpt{} processes the user’s request and proceeds to generate the corresponding vectorized C code. It uses the GPT-4 model to generate code completion.

\subsection{Checksum-based Testing} 
\label{sec:design-csum}
To ensure the correctness of the vectorized code, \vecgpt{} uses checksum-based testing to evaluate the equivalence of the original unvectorized and newly vectorized C programs. This is achieved by initializing the input arrays randomly, executing the functions, and comparing the outputs of both functions.
The output produced by \vecgpt{} is proven not equivalent or is marked as \textit{ plausible} (possibly correct).

\begin{figure}[htbp]
\centering
\includegraphics[scale=0.45]{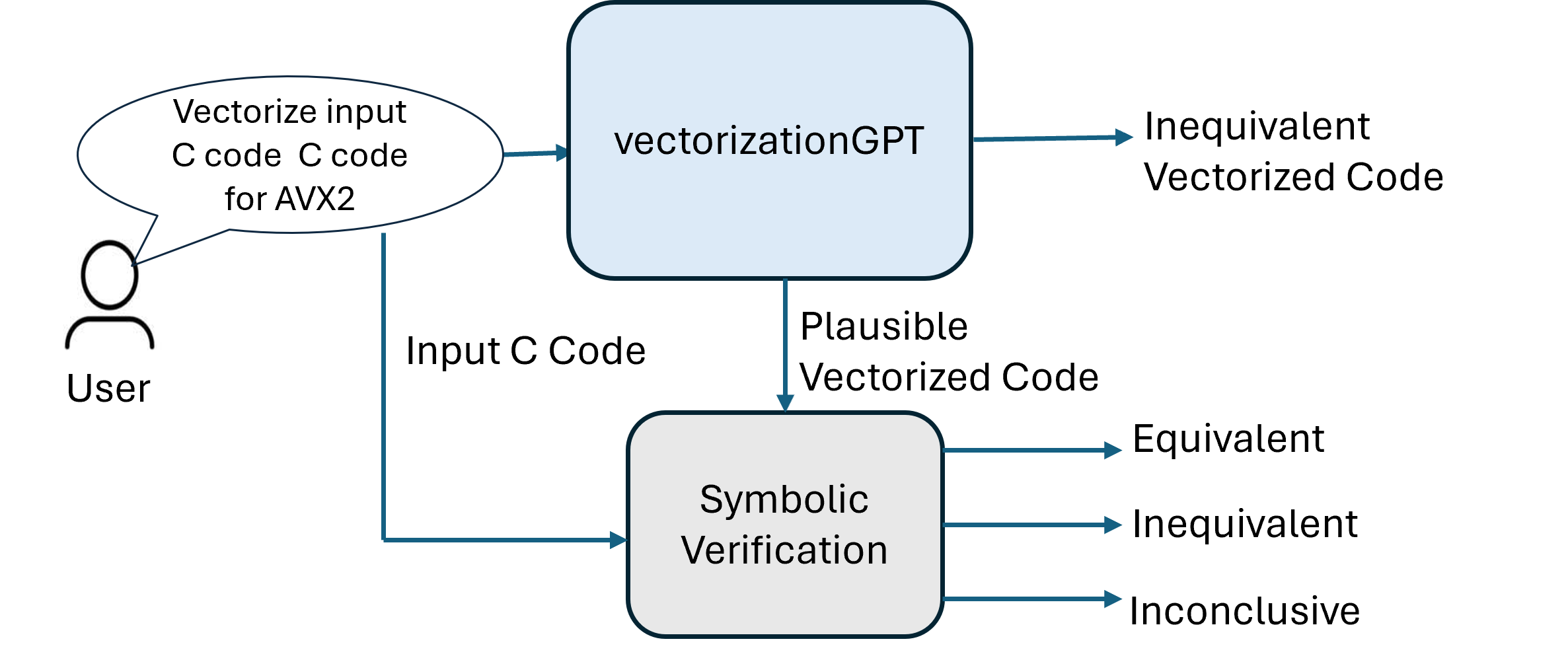}
\caption{High-level design overview of \vecgpt{}.}
\label{fig:main-design}
\end{figure}

Vectorized codes that appear to be plausible undergo further scrutiny through symbolic verification methods, enhancing the assurance of their correctness. VectorizationGPT’s symbolic verification relies on Alive2, a state-of-the-art translation validation tool for LLVM IR.
The high-level workflow of \vecgpt{} is depicted in Figure \ref{fig:main-design}.
%
%
\subsection{Multi-agents FSM}
\label{sec:agents}
In this section, we describe the design of \vecgpt{} that uses \textit{LLM agents} capable of intercommunication and excel in contextual information gathering for problem solving.
\subsubsection{Background: Large-language Model Agents}
LLM agents are intelligent entities that act as intermediaries between developers and large language models like GPT-4. These agents are designed to facilitate seamless communication with the underlying language model, enabling developers to harness its power effectively. We can think of them as the bridge that connects human intent with machine-generated text.
LLM Agents offer several advantages:
\begin{enumerate}
\item LLM agents abstract away the complexities of interacting directly with the raw language model. Instead of dealing with low-level API calls, developers can work with a simplified interface provided by the agent. This abstraction shields them from the intricacies of tokenization, context windows, and other technical details.
\item LLM agents handle context management efficiently. They maintain conversation history, context windows, and user prompts. This context awareness ensures coherent and relevant responses in multiple turns, making interactions more natural.
\end{enumerate}
Recent techniques have harnessed LLM-based agents and compiler feedback for code generation~\cite{fakhoury20243dgen}.
\subsubsection{\vecgpt{}}
\begin{figure*}
\centering
\includegraphics[scale=0.60]{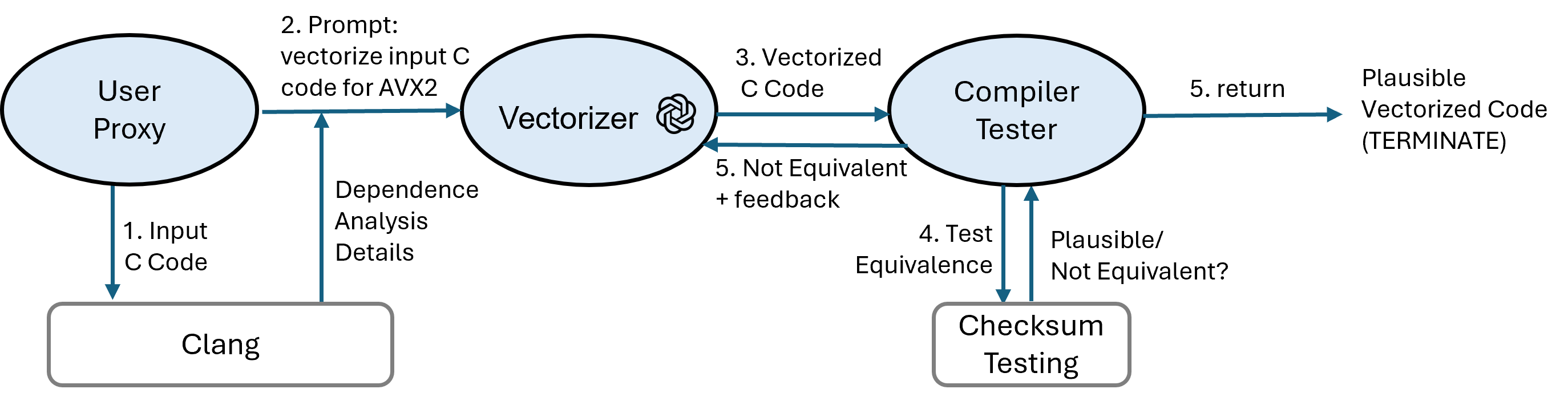}
\caption{Detailed design of \vecgpt{}.}
\label{fig:detail-design}
\end{figure*}

Our design harnesses these agents and orchestrates them using a finite state machine (FSM), allowing designers to easily specify transitions between agents.
This FSM is designed with two primary objectives in mind:
\begin{enumerate}
    \item reduce the number of LLM invocations required to generate correct vectorized code
    \item repair the incorrect vectorized code generated by LLM through a systematic feedback loop
\end{enumerate}
The workflow, illustrated in Figure \ref{fig:detail-design}, begins with a user proxy agent 
initiating a dialogue with a vectorizer assistant agent, providing code for vectorization and dependence analysis iinformation from the Clang compiler, highlighting why Clang cannot vectorize the loop. Dependence analysis identifies data dependence within the loop, i.e. if there is read-after-write, write-after-read data dependence across loop iterations. This analysis also determines loop carried dependence i.e. if an iteration depends on the result of the previous iteration.

The user instructs the vectorizer agent to eliminate the dependence for successful vectorization. Internally, the vectorizer agent consults the LLM and forwards both the original and vectorized code to the compiler tester assistant agent, which employs checksum-based testing to verify the plausibility of the vectorized code. If discrepancies arise, the compiler tester assistant provides feedback to the vectorizer agent, prompting code re-vectorization. This process repeats up to ten times or until a plausible solution is found.

Checksum-based testing involves setting a loop upper bound, initializing input arrays with random values, executing both code versions, and comparing output arrays. If the vectorized code fails to compile, it’s returned to the LLM for correction. Successful code passes through symbolic verification for soundness. In case of inequivalence, the tester agent prompts the vectorizer to resolve discrepancies, ensuring semantic equivalence between vectorized and unvectorized code.
The qualitative and quantitative results of the use of the multiagent FSM design are discussed in Section \ref{sec:fsm}.
%


\section{Equivalence Checking}
\label{sec:equivalence}
%
%
\newcommand{\inequivalent}{\textsc{Inequivalent}}
\newcommand{\plausible}{\textsc{Plausible}}
\newcommand{\equivalent}{\textsc{Equivalent}}
\newcommand{\inconclusive}{\textsc{Inconclusive}}
\newcommand{\checkEquivalence}{\textsc{checkEquivalence}}
\newcommand{\checksum}{\textsc{checksumTesting}}
\newcommand{\aliveUnroll}{\textsc{checkWithAlive2Unroll}}
\newcommand{\cUnroll}{\textsc{checkWithCUnroll}}
\newcommand{\caseSplitting}{\textsc{checkWithSpatialSplitting}}
\newcommand{\result}{\textit{result}}
%
%
\begin{algorithm}
\caption{Algorithm for Formal Equivalence Checking}
\label{algo:sym-verify}
\small{
\begin{algorithmic}[1]
\Procedure{\checkEquivalence}{S, V}
\State \result{} $\leftarrow$ {\checksum(S, V)}
\If {\result{} = \inequivalent}
\State \Return \result
\EndIf
\If {\result{} = \plausible}
    \State \result{} $\leftarrow$ {\aliveUnroll(S, V)}
    \If {\result{} $\neq$ \inconclusive}
         \State \Return \result
    \EndIf
    \State \result $\leftarrow$ {\cUnroll(S, V)}
    \If {\result{} $\neq$ \inconclusive}
        \State \Return \result
    \EndIf
    \State \result{} $\leftarrow$ {\caseSplitting(S, V)}
\EndIf
\State \Return \result
\EndProcedure
\end{algorithmic}
}
\end{algorithm}
\newcommand{\srcProg}[1]{\textit{S}}
\newcommand{\vecProg}[1]{\textit{V}}
%
In this section, we describe our approach to formally verify the correctness of vectorized code \vecProg{} produced by \vecgpt{} by checking them for equivalence against the input scalar code \srcProg{}. 
Algorithm~\ref{algo:sym-verify} describes the overall approach. 
We assume that the checksum based testing (\checksum{}, described in Section~\ref{sec:design-csum}) marks the pair as \plausible{} (i.e., is unable to distinguish between the two programs using concrete test cases).
\subsection{Symbolic Verification with Alive2}
\label{sec:no-unroll}
%
For the symbolic verification, we leverage Alive2, an automatic symbolic translator validation tool for the LLVM compiler's intermediate representation (IR) for checking equivalence. 
It is designed to ensure that the LLVM compiler's transformations do not introduce bugs by checking that the optimized target program {\it refines} the input program.
Alive2 provides a comprehensive modeling of {\it undefined behavior} (UB) in LLVM IR, a critical aspect since LLVM's optimizers often exploit UB. 
Alive2 has contributed to clarifying ambiguities in LLVM's IR specification and has led to numerous bug fixes and patches.
Using Alive2 also allows us to avoid defining semantics for vectorization intrinsics, such as \texttt{\_mm256\_blendv\_epi8}; Clang is responsible for emitting the proper LLVM IR for such intrinsics, which Alive2 then encodes into the logic of Satisfiability Modulo Theories (SMT-LIB~\cite{smt-lib-cite}) to be verified by Z3~\cite{de2008z3}.

Alive2 uses a bounded formal verification, where loops are unrolled a fixed number of times. 
This has the limitation that bugs that manifest only in a large number of unrolling may be missed by Alive2.
Unrolling loops increases the complexity of the verification query sent to the SMT solvers, resulting in the verifier returning \inconclusive{} results in the form of timeout or memoryout. 

In the next paragraphs, we describe the different challenges encountered to translate the vectorization correctness problem into a bounded translation validation problem (the method \aliveUnroll{} in Algorithm~\ref{algo:sym-verify}).
For this paper, we assume only \texttt{for} loops, where we represent the loops in scalar and vector programs in the canonical form:
$\texttt{for (i = start1; i < end1; i += step1) srcbody}$, and $\texttt{for (i = start2; i < end2; i += step2) vecbody}$; it is easy to extend to other ending conditions like $\texttt{i <= end}$ or $\texttt{i != end}$, or decrementing iterator such as $\texttt{i -= step}$).
We have several examples where $\texttt{step1}$ is not 1 and cases where either of both of $\texttt{end1}$ and $\texttt{end2}$ are complex expressions (see the target program in Fig~\ref{fig:motivation-perf}). 
Alive2 allows unrolling the loops in the source and target programs before comparing two loop-free programs for equivalence. 
%
\paragraph{Loop alignment}
For the vectorized program, a single loop iteration performs updates that are spread over several loop iterations in the scalar program.
This implies that the two loop-free programs created for the source and the vectorized program need to perform the same set of updates on both sides.
We also fix the unroll factor for the vectorized program to be 1. 
To compute the unrolling factor for the source program, we compute the least common multiple $v$ of $\texttt{step1}$ and $\texttt{step2}$, and unroll source for $m \doteq v/\texttt{step1}$ times. 
For simplifying the formal verification, we assume that the vectorized code does not require an epilogue scalar loop as the number of iterations of the scalar program is a multiple of the vectorization width. 
We ensure this by adding an assumption $\texttt{(end1 - start1)} \ \%  \ m == 0$ at the LLVM level through an ``assume'' instruction, where $\%$ denotes the modulus operator.
For the example in Fig~\ref{fig:motivation-perf}, where $m = 8$, we add the constraint $\texttt{assume (n - 1 - 0) \% 8 == 0}$.
We use a simple static analysis of the C source code and z3 SMT solver to obtain this bound. 
Our analysis currently does not handle cases where $\texttt{step1}$ is not a constant literal, or where the target introduces different number of loops compared to the source program.
%
%
\begin{figure*}
{\small
\begin{minipage}[t]{0.4\textwidth}
\begin{lstlisting}[language=C]
void s124(int *a, int *b, int *c, int *d, int *e, int n) {
  int j;
  j = -1;
  for (int i = 0; i < n; i++) {
    if (b[i] > 0) {
      j++;
      a[j] = b[i] + d[i] * e[i];
    } else {
      j++;
      a[j] = c[i] + d[i] * e[i];
    }
  }
}
\end{lstlisting}
(a) Unvectorized C code
\end{minipage}~
\hspace{0.032\textwidth}
\begin{minipage}[t]{0.6\textwidth}
\begin{lstlisting}[language=C]
#include <immintrin.h>

void s124(int *a, int *b, int *c, int *d, int *e, int n) {
  int j = 0;
  __m256i increment = _mm256_set1_epi32(1);
  __m256i zero = _mm256_setzero_si256();
  for (int i = 0; i < n; i+=8) {
    __m256i vbi = _mm256_loadu_si256((__m256i *)&b[i]);
    __m256i vci = _mm256_loadu_si256((__m256i *)&c[i]);
    __m256i vdi = _mm256_loadu_si256((__m256i *)&d[i]);
    __m256i vei = _mm256_loadu_si256((__m256i *)&e[i]);
    __m256i vprod = _mm256_mullo_epi32(vdi, vei);
    __m256i vsum_b = _mm256_add_epi32(vbi, vprod);
    __m256i vsum_c = _mm256_add_epi32(vci, vprod);
    __m256i vmask = _mm256_cmpgt_epi32(vbi, zero);

    __m256i va = _mm256_blendv_epi8(vsum_c, vsum_b, vmask);
    _mm256_storeu_si256((__m256i *)&a[j], va);
    j+=8;
  }
}
\end{lstlisting}
(b) Vectorized Code for \texttt{s124}.
\end{minipage}
}
\caption{Motivating example for Symbolic Verification. In vectorized code (b), \texttt{\_\_mm256\_blendv\_epi8} blends bytes from two $256$-bit integer vectors based on the most significant bit of each byte in a mask vector.}
\label{fig:motivation-verification}
\end{figure*}
%
\paragraph{Nested loops}
For nested loops, we have observed that only the inner loop needs to be vectorized, keeping the outer loop structure similar to the source program. 
Therefore, for nested loops, we check to make sure that the outer loops are syntactically identical and only align the inner loops. 
Given the identical outer loops, we only check the equivalence of the inner loops for an arbitrary iteration 
of the outer loop. 
%
\paragraph{Establishing non-aliasing}
To establish the correctness of vectorization one may need to assume some precondition about non-aliasing of the parameters. 
For example, consider the pair of source and vectorized programs, where the source program is  $\texttt{for (i = 0; i < n; i++) a[i] = b[i] + 1;}$ and the target program is:
\begin{footnotesize}
\begin{verbatim}
for (i = 0; i < n; i += 8) {
    __m256i b_vec = _mm256_loadu_si256((__m256i *)(b + i));
    __m256i one_vec = _mm256_set1_epi32(1);
    __m256i result_vec = _mm256_add_epi32(b_vec, one_vec);
    _mm256_storeu_si256((__m256i *)(a + i), result_vec);
}
\end{verbatim}
\end{footnotesize}
Since the addresses of the two arrays $\texttt{a}$ and $\texttt{b}$ are pointers, Alive2 may consider them to be aliased.
If we send this pair to Alive2, it finds a counterexample to equivalence, when $\texttt{b}$ aliases with the address of $\texttt{a-1}$ (i.e., $\texttt{\&b[1]}$ is the same pointer as $\texttt{\&a[0]}$).
For such an input, then source program effectively reduces to $\texttt{a[i] = a[i-1] + 1}$ which is not equivalent to the vectorized program, and perhaps not even vectorizable.

To tackle this issue, we ensure that the input arrays are allocated in different memory regions so that addresses of two different arrays never alias. 
To communicate this to Alive2, we define pointer $\texttt{a}$ and $\texttt{b}$ as external arrays (instead of function parameter); Alive2  models them as arrays allocated on different memory regions.
\subsubsection*{Example}
Figure \ref{fig:motivation-verification} shows a motivating example for which checksum based testing $\checksum$ found the vectorized code to be possibly equivalent $\plausible$, but Alive2 based symbolic verification identifies them as \inequivalent.
Figure \ref{fig:motivation-verification}(a) shows the scalar code, which conditionally loads from \texttt{c} in the \texttt{else} condition. 
All other arrays are read in both branches. 
However, the vector code in Figure \ref{fig:motivation-verification}(b) loads eight elements from \texttt{c} unconditionally.
Such a load may trigger undefined behaviour (specifically in the operation \texttt{\_mm256\_add\_epi32}), which Alive2 finds symbolically.
However, this bug is missed by checksum testing because such undefined behaviors may not exhibit in all runs.
Therefore, symbolic verification is an additional strategy we employ to increase confidence that codes produced by \vecgpt{} are correct.

With the techniques described in this section, we manage to show a few more examples \inequivalent{} using symbolic formal verification, and prove \equivalent{} under a bounded unrolling of loops.
However, the need to unroll loops often makes Alive2 return \inconclusive{} due to timeouts. 
In the next few subsections, we outline a few domain-specific optimizations to scale the symbolic verification.
\subsection{C-level Unrolling}
\label{sec:c-unroll}
Given that we have restricted the verification to cases where the number of iterations of the loops in the source scalar program to be a multiple of vectorization width, we can further simplify the verification condition.
For example, if the target program has a vectorization width of 8 (e.g., Fig~\ref{fig:motivation-perf}), then we know that the loop in the scalar program will execute multiples of 8 (including 0). 
This means that we can safely skip the check for loop termination in the source program $\texttt{i < end1}$ when $\texttt{i}$ is not a multiple of 8.
We achieve this by using an unrolling at the level  of scalar program at the C-level by a preprocessing step before generating the LLVM (instead of Alive2 performing unrolling for the LLVM program).

To unroll the source program for $v$ iterations, we replace the \texttt{for} loop as follow: First we initialize $\texttt{i}$ as $\texttt{i = start1}$, then unroll the loop body for $v$ times where we add $\texttt{i += step1}$ to the end of each loop body. We perform program analysis and slightly modify the loop body to ensure the unrolled program can compile as follows. $(1)$ We replace "break" statement with "return"; $(2)$ We replace  each "goto" label with a unique label to avoid using repeated "goto" labels across different iterations; finally, $(3)$ we remove duplicated variable declarations.

We also extend the unrolling at the C-level to nested loops. 
For a nested loop, we unroll only the inner loop. We check whether the outer loop remains exactly the same in source and target, then choose one iteration to verify. We choose an arbitrary iteration of the outer loop by elevating the loop iterator to a parameter value (e.g., if the outer loop is {\texttt for (i = 0; i < n; i++)}, we remove the {\texttt for} statement and elevate loop iterator {\texttt i} to be a function parameter).

We observe this simple transformation allows the symbolic verification to be scalable and terminate with a non \inconclusive{} result. 
We call this method \cUnroll{} in Algorithm~\ref{algo:sym-verify}, and invoke it if the default Alive2 algorithm \aliveUnroll{} does not succeed. 
\vspace{-2mm}
\subsection{Spatial Case Splitting}
\label{sec:slicing}
For several examples without any \textit{loop-carried dependencies}, one can decompose  the equivalence checking of an entire array into checking for the equivalence of each index of the array separately. 
This has the potential to simplify the verification complexity for the SMT solver, at the cost of making multiple verification queries. 
That is, to check that two arrays \texttt{a[0:$k$]} and \texttt{a'[0:$k$]} are equal for some fixed $k$, one can check $k$ equivalence queries each checking the equivalence of \texttt{a[$j$]} and \texttt{a'[$j$]} for each $j \in [0,k]$.
Moreover, for the $j^{th}$ query, one can only consider the $j^{th}$ index of the other arrays it reads from --- abstracting the content of all the arrays except at index $j$.
Further, in the absence of any loop-carried dependency and ensuring that iteration $j$ (in scalar program) updates $j^{th}$ entry of output arrays reading only from the $j^{th}$ entries of other arrays, we only retain a \textit{single loop iteration} of the scalar program that updates a fixed $j^{th}$ index. 
To achieve this, at entry we copy the content of each output array (say $\texttt{a}$) into a local array (say $\texttt{a1}$), and only update the $j^{th}$ index of output array with the value at the $j^{th}$ index in the  updated local array before exit (i.e., $\texttt{a[j] = a1[j]}$). 
The target program, similarly performs the vector operations on the local array (say \texttt{a1}) and also updates the output arrays at a single index $j$.
Since we only unroll the vectorized loop once, we generate $m$ equivalence queries, one for each value of $j \in [0, m)$, where $m$ is the vectorization width.
We term this  as \caseSplitting{} ({\it spatial case splitting}) in Algorithm~\ref{algo:sym-verify}.

To implement this optimization, we implement a (semi-automated at present) technique to conservatively check that the scalar and the vectorized programs do not have an loop-carry dependencies. 
In other words, each loop can be processed without any dependence on values from other iterations. 
Our loop-carry dependency at the C-level checks for the following: $(1)$  the source program accesses only the $i^{th}$ element in {\it every} array at each $i^{th}$ iteration, and the target accesses only vectors starting at the $i$th element (e.g., {\texttt \_m256\_loadu(\&a[i])}); and $(2)$ both programs do not update any scalar value across loop iteration. 
Our analysis is syntactic and conservative --- it fails to qualify the following program that has no loop-carry dependency for the loop-body
$\texttt{a[i] = a[i+1] + 1}$. 
For example, a semantically equivalent program that copies the original content of $\texttt{a}$ into an array $\texttt{tmp}$ ($\texttt{tmp[0:n-1] = a[1:n]}$) and rewrites the body as 
$\texttt{a[i] = tmp[i] + 1}$ does not have any loop-carry dependencies.

\section{Results}
\label{sec:results}
We design our experiments to address the following research questions.
\begin{itemize}
    \item \textbf{RQ1:} How well can GPT4 vectorize the code on its own?
    \item \textbf{RQ2:} Can the vectorized code be formally verified for equivalence?
     \item \textbf{RQ3:} Is the vectorized code performant?
    \item \textbf{RQ4:} Does the AI-agents help improve the quality of generated code?
\end{itemize}
\paragraph{Dataset.}
We used the Test Suite
for Vectorizing Compilers (TSVC)  benchmark to assess the vectorization capabilities of \vecgpt{}. This benchmark consists of $149$ for loops that operate on arrays of integer data type exclusively \cite{maleki2011evaluation}. In the dataset, these loops involve control flow, reduction, and data dependence. While the loops may have a fixed lower bound, the upper bound remains unknown. Additionally, the step count in the loops varies\textemdash either as a constant or a variable\textemdash across different tests. In our evaluation, each for loop is treated as an individual test program.
\paragraph{Experimental setup.}
The \vecgpt{} tool presented in this paper uses GPT-4 model at its core
to generate vectorized code. We do not make any fine-tuning adjustments to the
model and use it as it is. We configure GPT-4 model with a temperature set to
\textit{1.0} to enable more diversity and creativity in the responses. The API version
is set to \textit{2023-08-01-preview}.
For performance evaluation of LLM-generated vectorized code against state-of-the-art
compilers, we use \textit{GCC-10.5.0, Clang-19.0.0, ICC-2021.10.0} compilers. The details of
the compiler flags to compile unvectorized and vectorized programs are listed in
Table \ref{tab:flags}.
We run performance experiments on an eight-core \textit{Intel i7-8650U} CPU with \textit{16GiB} RAM and \textit{AVX2} target.
%
\begin{table*}
\caption{Compiler Optimization Flags and Version Details.}
\begin{center}
\scriptsize{
\setlength{\tabcolsep}{3pt}
{\begin{tabular}{l|l|l|l}
 \toprule
 Compiler & Version & Unvectorized & Vectorized \tabularnewline
 \midrule
  GCC & 10.5.0 & -O3 -mavx2 -lm  & -W -O3 -mavx2 -lm -ftree-vectorizer-verbose=3   \tabularnewline
       &  & & -ftree-vectorize -fopt-info-vec-optimized \tabularnewline
  Clang & 19.0.0 & -O3 -mavx2 -lm -fno-tree-vectorize & -O3 -mavx2 -fstrict-aliasing -fvectorize  \tabularnewline
        & & & -fslp-vectorize-aggressive -Rpass-analysis=loop-vectorize -lm \tabularnewline
  ICC & 2021.10.0 & -restrict -std=c99  -O3 -ip -no-vec & -restrict -std=c99  -O3 -ip -vec -xAVX2 \tabularnewline
 \bottomrule
\end{tabular}}
}
\end{center}
\label{tab:flags}
\end{table*}
%
\begin{table}[h]
\vspace{-2mm}
\caption{Evaluation of vectorized code using Checksum-based testing.}
\begin{center}
\setlength{\tabcolsep}{3pt}
{\begin{tabular}{lrrr}
 \toprule
 Parameters & k=1 & k=10 & k=100  \tabularnewline
 \midrule
  Plausible     & 72 & 107 & 125   \tabularnewline
  Not equivalent & 62 & 40  & 24    \tabularnewline
  Cannot compile & 15 & 2   & 0    \tabularnewline
 \bottomrule
\end{tabular}}
\end{center}
\label{tab:checksum}
\end{table}
%
\subsection{[RQ1] Understanding LLM's capability to generate vectorized code}
\label{sec:results-codegen}
%
\subsubsection{Results from Checksum-based Testing}
\label{sec:results-csum}
We conducted an experiment to evaluate the correctness of vectorized C functions generated by LLM. The results of this experiment are summarized in Table \ref{tab:checksum}.
We ran this experiment for $149$ tests from the TSVC benchmark for varying number of code completions.
The column labeled $k=1$ represent one code completion. In other words,
we prompt LLM to generate one output
(vectorized program) for each
input (scalar) program. We find that
$72$ tests are plausible.
To explore further, we increase the number of
code completions to $10$ and we find that $107$ test programs contain at least one plausible solution. To clarify, we consider a test
equivalent if we identify at least one pair of
unvectorized and vectorized functions with matching 
checksum results. 
However, if all ten vectorized functions 
generated by GPT-4 fail to be equivalent to the unvectorized 
function using Checksum-based testing (including cases where any compilation errors occur), we classify 
the test as not equivalent. 
If all the candidate solutions for an example fail the compiler, then we mark them under ``Cannot compile'' row. 
The $k=10$ column in 
Table \ref{tab:checksum} indicates that $107$ tests contain at least one plausible solution.
In order to reduce the number of inequivalent tests
in a given dataset of $149$ test programs, we continue to increase the
number of code completions to $100$. We found that $125$ tests
have at least one plausible vectorized program as per the checksum-based testing criterion.
%
%
\subsubsection{Measuring Success Metric, pass@k}
\newcommand{\passatk}{\textit{pass@k}}
To evaluate the quality of an LLM for the task of code generation (from natural language), while accounting for the statistical non-deterministic nature of LLMs, Chen et al.~\cite{chen2021evaluating} introduced the \passatk{} metric.  
Assuming the presence of a set of validation tests $T$, a code suggestion is considered "correct" if it satisfies all the tests in $T$.
Given a sufficiently large number of sample size $n$, they defined \passatk{} as the expected mean for a sample of size $k$ ($k$ usually smaller than $n$) to contain at least one correct solution. 
In our case, we adapt the metric to the case when a code suggestion is labeled \plausible{} by the Checksum-based testing. 

\begin{figure}[htb]
\centering
\includegraphics[scale=0.35]{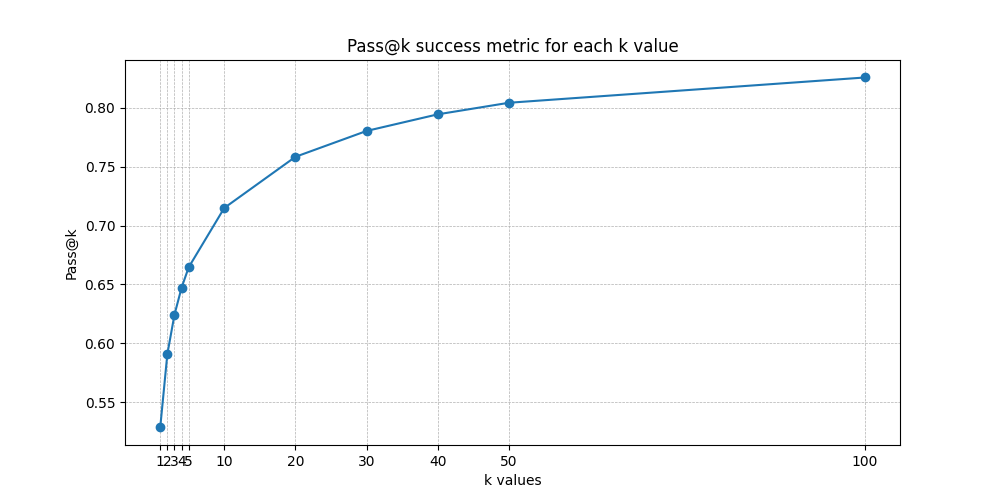}
\caption{Effectiveness of \vecgpt{} at generating correct vectorized code at different levels of $k$.}
\label{fig:passk}
\end{figure}

For each test program in the TSVC benchmark, we compute the `pass@k' value for different values of k (for example, 1, 2, 3, 4, 5, 10, 20, 30, 40, 50, 100) as shown on the x-axis of the chart in Figure \ref{fig:passk}.
At the end, we compute the average `pass@k' value over $149$ tests in the dataset for each `k'.

In Figure \ref{fig:passk}, the pass@k metric shows a steep increase as $k$ values rise from $1$ to $20$, indicating a rapid improvement in the \vecgpt{}'s performance with a small number of attempts.
The growth in pass@k begins to slow beyond $k = 20$, suggesting diminishing returns on the \vecgpt{}'s ability to generate additional correct vectorized code.
As $k$ approaches $50$, the pass@k metric reaches near-maximum, indicating that increasing $k$ beyond this point yields minimal improvement in correct vectorized code generation.
%
\subsubsection{Qualitative Analysis of LLM's Shortcomings}
\label{sec:results-failed}
In this section, we present a discussion about which codes \vecgpt{} could not vectorize.
In total, \vecgpt{} failed to vectorize $24$  tests.
Dependence is the main shared attribute among failed tests; for example, many tests contained loop-carried dependences.
By providing a simple dependence analysis through Clang feedback and input output examples through checksum-based testing, \vecgpt{} was able to succeed in more cases, as shown in Section \ref{sec:fsm}.
Despite this, \vecgpt{} still struggled with dependence relationships. In the following, we explain the two most common difficulties we have observed.

\textbf{\textit{One-time Dependence.}} 
Some loops contained a possible loop-carried dependence that could be satisfied at most once during execution. A common solution to these cases is loop \textit{splitting}, or \textit{distribution}, which partitions the loop domain in order to break up the dependence. We could not consistently get \vecgpt{} to perform the required transformations to vectorize such codes.

\textbf{\textit{Unsafe Hoisting.}}
Hoisting instructions out of a loop body is not always a safe transformation. In several tests, hoisting instructions resulted in easier to vectorize loops, but altered the code semantics. Due to poor dependence analysis, \vecgpt{} performed several unsafe hoists that could not be successfully repaired.
%
%

%
%
\subsection{[RQ2] Can LLM-generated vectorized code be formally verified?}
\label{sec:results-equivalence}
In this section, we assess the formal verifiability of the $125$ tests proven plausible using checksum-based testing. Initially, these $125$ tests are categorized as inconclusive, pending formal verification of their correctness.

Table \ref{tab:results-1} presents a summary of the results from various equivalence checking methodologies. The symbolic verification employing the out-of-the-box Alive2 unroll technique, described as \aliveUnroll{} in Section \ref{sec:no-unroll}, proves the correctness of vectorization for $26$ tests, leaving $82$ tests  inconclusive  due to multiple factors such as timeouts, out of memory, and unrecognized AVX2 intrinsics not yet encoded in Alive2.

Subsequently, we leverage the unrolling at the C program level for the  unvectorized  program (\cUnroll{} in Section \ref{sec:c-unroll}) within the subset of $82$ test programs that remained unproven by the preceding technique. This optimization enabled us to conclusively verify $18$ as non-equivalent, and  $28$ tests as equivalent (modulo loop unrolling). 
Notably, this method significantly reduced the incidence of timeouts by simplifying the queries sent to the SMT solver.

Finally, we applied the spatial-case splitting technique (\caseSplitting in Section~\ref{sec:slicing}) to the remaining dataset of $36$ tests. 
Unfortunately, a majority of the programs were filtered away due to our conservative static analysis to ensure lack of loop-carried dependencies.
After this selective filtering, we were left with five test programs, of which three were validated as equivalent and two as non-equivalent.

In conclusion, we can formally verify $57$ test programs as correct, refute $61$ tests, leaving 
$31$ as inconclusive.
Of these, our domain-specific
optimizations contributed to helping Alive2 refute 20 more vectorizations incorrect, and  verify  31 additional examples.

\begin{table}
\caption{Evaluation of Vectorized code using equivalence checking techniques. "Total" is total number of test programs, "Equiv", "Not Equiv" and "Inconcl" refer to \equivalent{}, \inequivalent{} and \inconclusive{} respectively.}
\begin{center}
{
\small
{\begin{tabular}{lrrrr}
 \toprule
 Techniques & Total & Equiv & Not Equiv & Inconcl \tabularnewline
 \midrule
  Checksum  & 149  & 0 & 24 & 125 \tabularnewline
  Alive2  & 125  & 26 & 17 & 82 \tabularnewline
  C-Unroll & 82 & 28  & 18  & 36  \tabularnewline
  Splitting & 36 &  3  &  2  & 31 \tabularnewline
\hline
  All &  149 & 57  & 61  & 31  \tabularnewline
 \bottomrule
\end{tabular}}
}
\end{center}
\label{tab:results-1}
\end{table}
%
%
\subsection{[RQ3] Is LLM-vectorized code faster?}
\label{sec:results-perf}
\begin{figure*}
\centering
\includegraphics[width=\textwidth]{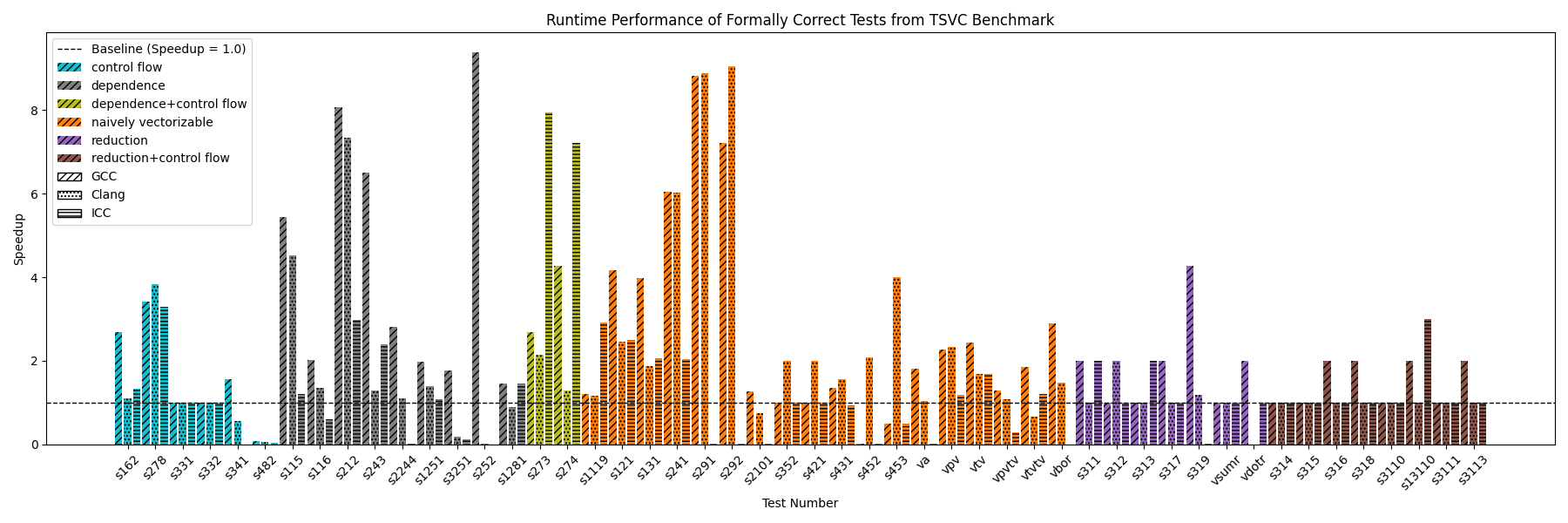}
\caption{Runtime performance speedup for formally correct test programs from TSVC benchmark.}
\vspace{-1mm}
\label{fig:perf-all}
\end{figure*}

In this section, we compute the runtime performance speedup
for $57$ equivalent tests proven by equivalence checking.
We found that LLMs are capable of producing high-performance vectorized code with run-time speedup ranging from $1.1x$ to $9.4x$ as compared to the state-of-the-art compilers.
We classify each test into one of six categories, each represented by a different color in the figure.
Each test is associated with three different bars\textemdash slanted line, dots, and horizontal line\textemdash to show the performance of GPT4 against GCC, Clang, and ICC, respectively 
as shown in Figure \ref{fig:perf-all}.
The baseline speedup is set at $1.0$, representing the scenario where state-of-the-art compiler and GPT-4 exhibit equivalent performance. Any speedup value below $1.0$ indicates a slowdown in GPT-4's performance relative to the other compiler.
The following paragraphs provide an analysis of each category.
\paragraph{\textit{Control Flow.}}
GPT4 shows the highest performance benefits in test \textit{s278} (for loop shown below), which contains complicated control flow with \texttt{goto} statements that requires selection instructions to vectorize efficiently. 
Otherwise, general-purpose compilers can vectorize many codes with control flow by using if-conversion; therefore, other tests in this category show no speedup, or some slowdown, where the compiler does a better job of vectorization.
\begin{footnotesize}
\begin{verbatim}
for (int i = 0; i < n; i++) {
  if (a[i] > 0) {
    goto L20;
  }
  b[i] = -b[i] + d[i] * e[i];
  goto L30;
L20:
  c[i] = -c[i] + d[i] * e[i];
L30:
  a[i] = b[i] + c[i] * d[i];
}
\end{verbatim}
\end{footnotesize}
\paragraph{\textit{Dependence.}}
Unlike control flow, which is usually straightforwardly vectorized by compilers, loops with dependence can disable vectorization entirely due to imprecise dependence analysis. 
GPT4 therefore shows several significant speedups in this category. 
In most cases, GPT4 has the largest speedups over GCC and Clang, and smallest speedups (or slowdowns) over ICC. 
Because ICC performs a sophisticated dependence analysis that is tightly integrated with the loop vectorizer, it tends to produce fast vector code, even with dependences.
However, GCC and Clang often disable vectorization entirely.
\paragraph{\textit{Dependence+Control Flow.}} 
A mixture of dependences and control flow are difficult for compilers to vectorize; therefore, GPT4's more aggressive vectorization strategy always attains speedups in this category.
For example, we include \textit{s274} below:
\begin{footnotesize}
\begin{verbatim}
for (int i = 0; i < n; i++) {
  a[i] = c[i] + e[i] * d[i];
  if (a[i] > 0) 
    b[i] = a[i] + b[i];
  else
    a[i] = d[i] * e[i];
}
\end{verbatim}
\end{footnotesize}
In this case, control flow is mixed with loop-carried dependencies involving \texttt{a}.
\paragraph{\textit{Na\"ively Vectorizable.}} 
This category is a broad grouping of loops that are generally ``easy'' to vectorize for most compilers; they do not contain any complicated features like control flow or dependence relations.
Because compilers produce efficient vector code for such loops, the most number of slowdowns and negligible speedups for GPT4 are observed in this category.
Notable exceptions include tests \textit{s291} (shown below) and \textit{s292}, which although do not contain control flow or dependence, require transformations such as loop peeling for the best performance. ICC is able to achieve such performance automatically, while GCC and Clang cannot.
\begin{footnotesize}
\begin{verbatim}
int im1;
im1 = n-1;
for (int i = 0; i < n; i++) {
  a[i] = (b[i] + b[im1]) * 2;
  im1 = i;
}
\end{verbatim}
\end{footnotesize}
\paragraph{\textit{Reduction.}}
Unlike na\"ively vectorizable loops, reductions involve loop-carried dependence that requires special handling to vectorize. 
However, the reduction pattern is so common that compiler support for reductions is robust.
This is shown in Figure \ref{fig:perf-all} by either small speedups or slowdowns for GPT4.
For example, below we include for loop from test \textit{vsumr}:
\begin{footnotesize}
\begin{verbatim}
sum = 0.;
for (int i = 0; i < n; i++)
  sum += a[i];
\end{verbatim}
\end{footnotesize}
Such a straight-forward reduction pattern is very well supported by vectorization passes in general-purpose compilers.
\paragraph{\textit{Reduction+Control Flow.}}
Reductions mixed with control flow can be vectorized by applying if-conversion.
Therefore, this category is also amenable to traditional compiler vectorization techniques, and GPT4's speedups are less significant.
However, up to a $2\times$ speedup is still obtained over ICC.
%

%
%
\subsection{[RQ4] Evaluation of Multi-agent FSM}
\label{sec:fsm}
While addressing the initial research question delineated in Section \ref{sec:results-csum}, our findings indicate that $125$ tests have been validated as plausible. However, this required up to $100$ LLM invocations to ascertain an equivalent vectorized code for numerous test programs, which proved to be very expensive.
In this section, we aim to assess whether the implementation of a multiple agents finite state machine, in conjunction with auxiliary tools, can effectively reduce the frequency of LLM invocations. Additionally, we will explore the potential of multiple agents in facilitating the repair of non-equivalent vectorized programs through a feedback loop mechanism.
\subsubsection{Number of LLM Invocations}
In our study, using the checksum-based testing criterion of equivalence, we identified $72$ plausible tests when generating only code completions, as detailed in Table \ref{tab:checksum}.
Notably, we discovered $24$ new tests that were previously unsuccessful with a single code completion but are now successfully generated using multiple agents FSM. Consequently, the total number of plausible tests increased to $96$ when considering just one LLM invocation, compared to the initial $72$ tests. Furthermore, despite generating $100$ code completions, we encountered four test programs that did not have a plausible vectorization previously but now have a plausible vectorization with just a single code completion in multi-agents design. These findings suggest that the design involving multiple agents along with auxiliary tools holds promise for reducing the number of LLM calls and improving the quality of the generation.
%
\subsubsection{Qualitative Analysis of multiple agents towards repair}
\label{sec:results-repair}
In this section, we study the capability of multiple agents by taking one example in facilitating the repair of non-equivalent vectorized programs through a feedback loop mechanism. 
For the function \textit{s453}, the multi-agents FSM takes two attempts to successfully generate
vectorized code. The input scalar program is shown here.
\begin{footnotesize}
\begin{verbatim}
void s453(int *a, int *b, int n) {
    int s = 0;
    for (int i = 0; i < n; i++) {
        s += 2;
        a[i] = s * b[i];
    }
}
\end{verbatim}
\end{footnotesize}
\paragraph{First Attempt.} The vectorizer assistant agent starts by naively transforming scalar operation $s += 2$ to a SIMD operation. It initializes a vector $s\_vec$ to zero, which is used to store the scalar $s$ in vector form.
It sets up a vector $two\_vec$ with all elements as $2$, which is used to increment $s\_vec$ in each iteration.
Subsequently, it vectorizes all other operations and increments $s$ by $16$, corresponding to the simultaneous processing of eight elements.
Here is the vectorized code:
\begin{footnotesize}
\begin{verbatim}
void s453(int *a, int *b, int n) {
    __m256i s_vec = _mm256_set1_epi32(0);
    __m256i two_vec = _mm256_set1_epi32(2);
    __m256i s_increment = _mm256_set1_epi32(16);
    int i = 0;
    for (; i <= n - 8; i += 8) {
        s_vec = _mm256_add_epi32(s_vec, two_vec);
        __m256i b_vec = _mm256_loadu_si256((__m256i*)&b[i]);
        __m256i a_vec = _mm256_mullo_epi32(s_vec, b_vec);
        _mm256_storeu_si256((__m256i*)&a[i], a_vec);
        s_vec = _mm256_add_epi32(s_vec, s_increment);
    }
}
\end{verbatim}
\end{footnotesize}
When compiler tester agent tests the equivalence using checksum-based testing, it finds a discrepancy in the actual and expected output arrays. It sends the feedback to the vectorizer agent.
The vectorized code in the first attempt is not equivalent to the original
scalar code because it updates value of $s$ to $2$ in a vector slice
comprising of eight elements.
In contrast, the scalar code updates $s$ by $2$ in every iteration, which
means the values of $s$ should be updated to ${2, 4, 6, 8}$, and so on.
Let's see how vectorizer understands this issue and fixes it in the second attempt.
\paragraph{Second Attempt.} The vectorizer agent uses the input and output values of arrays
and attempts to vectorize again. This time, it vectorizes correctly by initializing
$s\_vec$ with a set of values as expected. There is no need to increment each value in $s\_vec$ by two anymore, but just by $16$ to operate on next slice of eight elements.
Here is the vectorized code:
\begin{footnotesize}
\begin{verbatim}
void s453(int *a, int *b, int n) {
    __m256i s_vec = _mm256_setr_epi32(2, 4, 6, 8, 10, 12, 14, 16);
    __m256i two_vec = _mm256_set1_epi32(16);
    int i = 0;
    for (; i <= n - 8; i += 8) {
        __m256i b_vec = _mm256_loadu_si256((__m256i*)&b[i]);
        __m256i a_vec = _mm256_mullo_epi32(s_vec, b_vec);
        _mm256_storeu_si256((__m256i*)&a[i], a_vec);
        s_vec = _mm256_add_epi32(s_vec, two_vec);
    }
}
\end{verbatim}
\end{footnotesize}
The compiler tester agent finds this solution to be correct.
The process terminates and the tester returns the correct vectorized code. This example is also formally verified and proven equivalent using Alive2 unroll technique.

We ran this experiment on $149$ tests in the TSVC benchmark and configured multi-agent FSM
to allow a maximum of ten attempts to solve the problem. We found correct vectorized
codes for the $92$ tests. In the evaluation of $92$ tests, agents required multiple iterations to resolve nine of them. The maximum number of attempts recorded to successfully address an issue within a single test was seven.
%
%

%

\section{Related Work}
\label{sec:related}
%
\textbf{Compiler auto-vectorization.}
Maleki et al. \cite{maleki2011evaluation}
studies the efficacy of compilers in vectorizing a synthetic benchmark and real applications, revealing that a significant portion of loops remain unvectorized, highlighting the complexity of auto-vectorization in the presence of intricate control flows.
Siso et al. \cite{siso2019evaluating}
assess compilers' vectorization capabilities by systematically withholding information that aids the auto-vectorization process, thereby providing a more realistic gauge of compilers' performance in practical scenarios.
Allen et al. \cite{allen1983conversion}
propose a systematic method to convert control dependence into data dependence, facilitates the application of data dependence-based program transformations. 
%
%

\textbf {ML-guided Compiler Optimizations.}
In recent years, the use of data-driven approaches to compiler optimizations has been explored as an alternative to traditional rule-based heuristics.
NeuroVectorizer\cite{haj2020neurovectorizer} utilizes an end-to-end deep reinforcement learning based approach for vectorization decisions. 
Mendis et al. \cite{mendis2019compiler} leverages imitation learning to replicate the decisions of an optimal Integer Linear Programming (ILP) model, generating vector code that is both efficient and functionally equivalent. 
Ashouri et al. \cite{ashouri2022mlgoperf} 
integrates a machine learning model with LLVM’s inlining process, and predicts the potential performance gains from inlining decisions. 
%
%
%
%
More recently, Cummins et al. \cite{cummins2023large} leverage fine-tuned LLMs to enhance compiler heuristics, aiming to optimize program execution without the need for extensive manual tuning. 
Grubisic et al. \cite{grubisic2024compiler} extend this line of work by exploring the use of real-time compiler feedback for LLMs to learn and improve subsequent compilation.
%
%

In contrast to all the prior uses of ML/LLM for compiler optimization, we use the emergent capabilities of LLMs such as GPT-4 to directly perform a source-to-source rewriting of entire loop bodies. 
Unlike traditional compiler optimizations, these transformations can be incorrect. 
Hence we propose employing testing and formal verification techniques to verify the correctness of these transformations. 

\textbf{Formal verification for compilers.}
In addition to Alive2, translation validation is implemented by a diverse range of existing tools~\cite{leroy2008formal,kasampalis2021language,lopes2021alive2,steppEqSat}.
Although some of these techniques offer the potential to perform unbounded verification (e.g., through equality saturation in Stepp et al.~\cite{steppEqSat}), they are not as automated and robust as Alive2 for the entire LLVM instruction set. 
%

\section{Conclusion}
\label{sec:conclusions}
In this paper, we developed \vecgpt{} to evaluate the capabilities of LLMs and AI-based agents towards generating correct and efficient loop vectorization optimizations leveraging compiler intrinsics. 
Our study highlights opportunities and challenges to automate  vectorizing programs through LLMs and formal verification. 
For future work, we plan to add feedback of failure and rootcause~\cite{lahiri-symdiff-rootcause} from the equivalece failure within \vecgpt{}, as well as extend the assurance of vectorized code by automating unbounded translation validation.
\section*{Acknowledgement}
The authors extend their sincere appreciation to Saeed Maleki for providing guidance and insight on vectorization. The authors also thank Sarah Fakhoury for her advice on the design of AI agents and Alive2 developers including Nuno Lopes and Zhengyang Liu for their clarifications and detailed review on the semantics of LLLVM IR.
%
\bibliographystyle{ACM-Reference-Format}
\bibliography{bib.bib}
\end{document}